\documentclass[aip,twocolumn]{revtex4}

\usepackage{amsmath,amssymb}
\usepackage{graphics,graphicx}

\usepackage[dvips]{epsfig}
\usepackage{wrapfig}
\usepackage{float}
\usepackage{multirow}

\newcommand{\rv}{{\bf r}}

\newcommand{\tv}{\hat{{\bf t}}}

\usepackage{color}
\newcommand\greg[1]{\bgroup\color{red}\bfseries{ [Greg: #1]}\egroup}
\newcommand\new[1]{\bgroup\color{blue}\bfseries{#1}\egroup}

\begin{document}

\title{Theory of Crosslinked Bundles of Helical Filaments:  Intrinsic Torques in Self-Limiting Biopolymer Assemblies}

\author{Claus Heussinger}

\affiliation{Georg-August-Universit\"at G\"ottingen, Institut f\"ur Theoretische Physik, Friedrich-Hund-Platz 1, 37077 G\"ottingen, Germany}
\affiliation{Max
  Planck Institute for Dynamics and Self-Organization, Bunsenstr. 10, 37073
  G\"ottingen, Germany} 
  \author{Gregory M. Grason}
\affiliation{Department of Polymer Science and Engineering, University of Massachusetts, Amherst, MA 01003, USA}
\begin{abstract}
  Inspired by the complex influence of the globular crosslinking proteins on the
  formation of 
  biofilament bundles in living organisms, we study and
  analyze a theoretical model for the structure and thermodynamics of bundles of
  helical filaments assembled in the presence of crosslinking molecules.  The
  helical structure of filaments, a universal feature of biopolymers such as
  filamentous actin, is shown to generically frustrate the geometry of
  crosslinking between the ``grooves'' of two neighboring filaments. We develop
  a coarse-grained model to investigate the interplay between the geometry of
  binding and mechanics of both linker and filament distortion, and we show that
  crosslinking in parallel bundles of helical filaments generates {\it intrinsic
    torques}, of the type that tend to wind bundle superhelically about its
  central axis.  Crosslinking mediates a non-linear competition between the
  preference for bundle twist and the size-dependent mechanical cost of filament
  bending, which in turn gives rise to feedback between the global twist of
  self-assembled bundles and their lateral size.  Finally, we demonstrate that
  above a critical density of bound crosslinkers, twisted bundles form with a
  thermodynamically preferred radius that, in turn, increases with a further
  increase in crosslinking bonds.  We identify the {\it stiffness} of
  crosslinking bonds as a key parameter governing the sensitivity of bundle
  structure and assembly to the availability and affinity of crosslinkers.
\end{abstract}
\maketitle

\section{Introduction}

Rope-like assemblies of filamentous biopolymers are important and common
structural elements in living organisms.  Fibers of cellulose and collagen
provide mechanical reinforcement of extra-cellular plant and animal
tissue~\cite{fratzl} while inside of eukaryotic cells, bundles of cytoskeletal
protein filaments -- microtubles, filamentous actin (f-actin) and intermediate
filaments -- are implicated in an outstanding array of physiological processes,
from cell division and adhesion to motility and mechanosensing~\cite{theriot}.
Understanding the physical mechanisms that underly the robust and high fidelity
assembly pathways of protein filaments is thus an important, outstanding
challenge with broad implications in biology.  Key questions revolve around the
role played by the myriad types of relatively compact, crosslinking proteins
that coassemble in parallel bundles and fibers of certain protein filaments, the
primary example of these being parallel actin bundles~\cite{pollard, bartles}.
Though the intrinsic properties of f-actin are largely conserved among different
cell types and species, the structural, mechanical and dynamic properties of
f-actin bundles are evidently quite modular.  For example, filopodial
bundles~\cite{faix, aratyn} that form at the periphery of motile cells are
loosely organized and highly dynamical structures of rather limited diameter
($\sim$ 100 nm), while f-actin bundles formed in mechanosensory appendages of
the cochlea and inner ear~\cite{tilney_saunders} are nearly crystalline in
cross-section and quite large by comparison ($\sim$ 1 ${\rm \mu}$m).  Recent
experimental studies of parallel actin bundles {\it in
  vitro}~\cite{shin_mahadevan, shin_mahadevan_pnas, claessens_nat_06,
  claessens_pnas_08, purdy, shin} demonstrate that to large extent the
mechanical and structural properties of bundle assemblies can be attributed to
the crosslinking proteins as well as their interactions with the bundled
filaments.

An important but unresolved question concerns the organization of crosslinks in
the bundle and how this organization reflects structural features of the
filaments themselves.  In particular, protein filaments are universally helical
in structure~\cite{squire}, and by virtue of the helical distribution of binding
sites, interactions mediated by proteins crosslinking neighboring filaments must
reflect the underlying chiral nature of the assembly.  Indeed, there is numerous
experimental evidence to show that crosslinking actin in parallel bundles
modifies the torsional state of constituent filaments from their native, unbound
geometry~\cite{derosier_censullo, tilney_derosier_mulroy, derosier_tilney,
  tilney_saunders, claessens_pnas_08, purdy, shin}.  The extent to which this
helical twist is transfered more globally to the structure of crosslinked actin
bundles -- as has been observed in superhelical assemblies of
fibrin~\cite{weisel} and collagen fibers~\cite{ottani} -- is presently unclear.

In this paper, we explore the fundamental interplay between the helical
structure of biological protein filaments and intrinsic torques generated by
self-organizing distributions of crosslinks between filaments in regular bundle
arrays.  Our task is to develop a generic theoretical model for the frustration
that arises between the regular in-plane organization of filaments in bundles
and the helical distribution of crosslinking sites along the filaments.  We seek
to understand how this frustration can be relieved at the expense of different
types of mechanical distortions of the filament assembly, in particular, a
global twist distortion of the parallel array.  A number of
theoretical~\cite{turner, grason_prl_07, grason_pre_09},
computational~\cite{hagan} and experimental~\cite{weisel} studies demonstrate
that chiral filament interactions which tend to twist bundled assemblies have
the important consequence of providing an intrinsic and thermodynamic limitation
to the lateral growth of bundles.  In the present study, we show that the
helical distribution of crosslinking sites on the filaments gives rise to a
tendency for neighboring filaments to twist in order to relieve the elastic cost
of distorting crosslinks.  In turn, the tendency to twist filaments
superhelically in the bundle leads to a mechanical cost for growing the bundle
diameter that for sufficiently weak crosslink affinities and sufficiently high
crosslink densities becomes finite in equilibrium.

Our approach to this problem is based on a generalization of the ``worm-like
bundle'' model of crosslinked filaments~\cite{bathe,heussinger, heussinger_pre}.
This semi-microscopic model treats filaments as semi-flexible polymers decorated
by discrete arrays of sites to which elastic crosslinks between neighboring
filaments bind.  This model has been successfully used to analyze the complex
mechanical response of crosslinked bundles to bend and twist deformations.
However, it does not allow to explore the consequences of the helical filament
structure for the bundle mechanical and thermodynamical properties.

In the present study, we consider a model where crosslink sites are located
along helical ``grooves'' on the filaments, which are characterized by bending
and torsional stiffness. Modeling the elastic cost of linker distortions, we show that crosslinking
between helical filaments leads to a energetic preference to align the opposing
grooves of crosslinked filaments.  Depending on the relative stiffness of the
linker and the filaments we find two different regimes: 1) a high-torque
``groove-locked'' regime, where stiff crosslinks force the grooves into
alignment, and 2) a low-torque ``groove-slip'' regime, in which crosslinks are
not stiff enough to enforce this alignment and cannot unwind the filaments from
their native state of twist. Though of a distinct microscopic origin, we find
that in the groove-locked regime linker-mediated interactions lead to a similar
elastic frustration as occurring in ``coiled-coil'' assemblies of
polypeptides~\cite{neukirch}.  Filament pairs may align opposing grooves by
either untwisting the pitch of grooves themselves {\it or} by winding helically
around one another with the appropriate pitch.

Based on the intrinsic and non-linear torques generated by crosslinking in
helical filament bundles, our model makes two important predictions.  First, we
predict that subject to an external torque self-assembled bundles of crosslinked
helical filaments exhibit non-linear torsional response that is highly-sensitive
to both the intrinsic helical geometry of filaments as well as the fraction of
bound crosslinks, $\rho$.  Second, we show that the competition between the
linker-generated torques and the mechanical energy of bending filaments in
superhelical bundles gives rise to the formation of self-limited bundles when
the crosslink fraction is larger than a critical value $\rho_c$, which is itself
determined by the ratio of bending cost of helically winding a pair of
  filaments to the torsional cost of unwinding the intrinsic twist of the
  helical grooves.  A primary conclusion of this study is that finite-diameter
bundles of helical filaments form preferentially when crosslinks are highly
resistant to in-plane shear distortions and filaments have a large torsional
stiffness relative to the bending modulus.

This article is organized as follows.  In Sec. II we introduce our model of
crosslinked helical filament assemblies and in Sec. III we derive the form of
the linker-mediated torsional energy of filament bundles.  In Sec. IV we
determine the dependence of bundle twist on bundle size, as well as the
torsional response of self-assembled bundles.  In Sec. V we predict the
thermodynamic behavior of crosslinked filament assemblies in terms of the
density and binding energy of crosslinks in the bundle.  Finally, we conclude
with a discussion of our results in the context of biological filament
assemblies.

\section{Model of Crosslinking in Helical Filament Bundles}

\subsection{Geometry of Crosslinked Helical Filament Bundles}

\begin{figure*}
  \center \includegraphics[width = 3.75in]{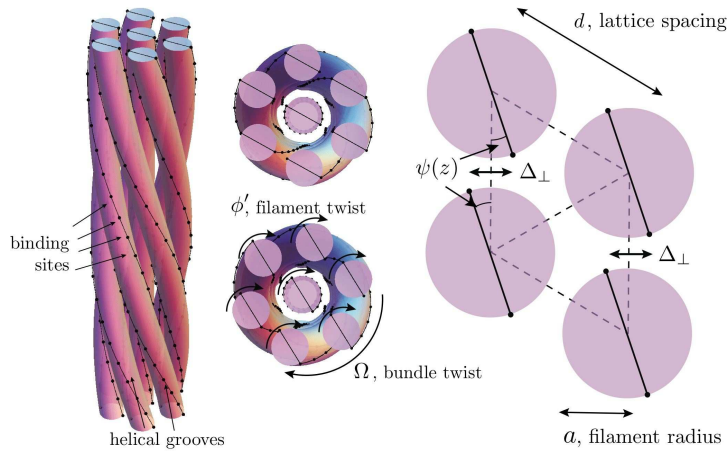}\caption{Illustration
    of the bundle geometry. Filaments are arranged on the sites of a
    two-dimensional hexagonal lattice. Discrete crosslink binding sites are
    regularly spaced along the filament backbone, respecting the double-helical
    structure of the filaments.  The binding sites on two neighboring filaments
    may be connected by crosslinkers. When bound, crosslink mediated torques act
    to align the binding sites on the two filaments. Alignment can be achieved,
    either by filament twist $\phi'$, where filaments are untwisted from their
    native helical state, or by bundle twist, where filaments superhelically
    wind around the center of the bundle. Misaligned binding sites imply an
    elastic cost for the crosslinkers, which we parametrize in terms of the
    angle difference $\psi$ with the optimal alignment and the associated
    ``in-plane'' shear deformation $\Delta_\perp\sim a\psi$.}
\label{fig: model}
\end{figure*}

To describe the interaction between crosslinking in parallel bundles and the
helical geometry of constituent filaments, we introduce the following
coarse-grained model, depicted schematically in Fig. \ref{fig: model}.  In the
cross-section (``in-plane''), the bundle is organized into a hexagonal array,
with the center-to-center spacing of neighbor filaments, $d$.  Crosslinkers bind
to neighboring filaments in parallel bundles, and reflecting the helical
symmetry of the filament, the ends of crosslinks are located at discrete points
on helical grooves on the filaments.  In the most general case, helical
filaments possess a range of groove geometries of differing helical symmetry.
For the purposes of the following analysis, we focus on the simple case where
binding sites are located on double-helical grooves, which are perfectly out of
phase ($180^\circ$ between grooves).  The crosslinking sites are linearly spaced
by a vertical separation, $\sigma^{-1}$, along the filament backbone direction,
and the pitch of each helical groove is defined to be $2 \pi / \omega_0$.

Albeit simpler, this double helical geometry is not unlike the two-start helical
structure of f-actin~\cite{holmes}, whose grooves rotate at a rate $\omega_0 = 2
\pi \sigma/ 13$ in the native state of twist and $\sigma^{-1} \simeq 5.7 \ {\rm
  nm}$ represents the vertical distance separating 2 actin monomers along the
same bi-helical groove.  While in actin filaments, the monomers on opposing
grooves are off-set vertically by $\sigma/2$, the present simplified model has
the advantage that in the lowest energy state, which allows for maximal crosslinking, all filaments
maintain the same axial ``orientation'', meaning that grooves in each vertical
plane are aligned along the same direction.  Certainly, the analysis may be
extended to address groove geometries that lead to non-trivial, inter-filament
correlations, as has been done in ref. \cite{shin_10}, but our purpose here is
to analyze the simplest model describing the interplay between helical geometry
of filaments and the inter- and intra-filament torques that arise in crosslinked
parallel bundles.

In parallel bundles, crosslinks bind selectively to pairs of sites which are
closest in separation (see Fig.~\ref{fig: bindingzone}), which in our model
occur when grooves on any neighbor filament ``cross'', and the orientation of
the groove at a given plane is perfectly aligned with the in-plane vector
separating the neighbors.  For perfectly straight filaments with native twist,
such a crossing occurs every $\pi/\omega_0$ along the backbone between any
neighbor pair, and over a distance $2 \pi /(6 \omega_0)$, the crossing rotates
from one of the six neighbors of a given filament to the next.  Here, we will
consider the case where the fraction, $\rho$, of occupied crosslinking sites in
the bulk of the bundle is fixed.  By the assumption of the minimal crosslink
stretching, crossing points are populated first, followed by sites closest to
the crossing point, and so on, until a ``binding zone'' representing the
$60^\circ$ wedge shared by any 2 neighboring filaments is occupied with the
appropriate number of crosslinks (see Fig.~\ref{fig: bindingzone}).

\subsection{Mechanics of Crosslinks and Twist Geometry of Filaments and Bundle}

To model the elastic cost of crosslinks that are poorly aligned due to the
helical distribution of sites we introduce the following simple crosslink
energy,
\begin{equation}
  \label{eq: elink}
  E_{link} = - \epsilon + \frac{k_\perp}{2} \Delta_\perp^2 .
\end{equation}
Here, $E_{link}$ is the energy of a single crosslinker in the bundle, and
$-\epsilon$ represents the ``bare'' energy gain for binding between 2 perfectly
aligned sites.  The second term represents the elastic cost of {\it shear}
distortions of crosslinks away from the perfectly aligned geometry perpendicular
to the backbone orientations.  Specifically, $\Delta_\perp$ represents the {\it
  in-plane shear} as shown in Fig. \ref{fig: model}.  This model has marked
similarities with a model for contact between hydrophobic residues on
$\alpha$-helical polypeptide chains studied in ref. \cite{wolgemuth_sun}, and
hence, our current model may also be applicable to the study of coiled-coil
bundles of $\alpha$-helices.

Depending on the location of the crosslink in the bundle, these distortions,
$\Delta_\perp$, are purely determined by the state of
twist of constituent filaments as well as the global twisting of the bundle
itself.  The twist of individual filaments is described by the angle, $\phi(z)$,
that describes the orientation of the groove in the $x-y$ plane of filament
packing.  Hence, for the case of native filament twist, $\phi ' = \omega_0$.  In
addition to the ``local'' twist of individual filaments, the bundle may twist as
a whole described by $\Omega$, the rate at which in-plane lattice directions
rotate around $z$ axis, the long axis of he bundle.  Notice that in this
description, $\phi'$ and $\Omega$ are decoupled by construction, meaning that
filaments may be twisted without the bundle experiencing twist ($\phi' \neq
\omega_0$ and $\Omega =0$) or the bundle may be twisted without distortion the
native symmetry of the constituent filaments ($\phi'=\omega_0$ and $\Omega \neq
0$)~\endnote{Strictly speaking, bundle twist and the rotation of helical grooves are coupled geometrically, since even for $\phi'=0$, when $\Omega \neq 0$ the writhe of helically bent filament backbones will lead to precession of groove orientation in the $x-y$ plane of linker binding at vertical height, $z$.  By treating rotations of bundle lattice and filament grooves independently, we neglect small, position dependent corrections to $\phi'$ needed to maintain constant groove orientation in the $x-y$ plane.  It can be shown than these corrections contribute to the bundle free energy density by a factor proportional to $C\Omega^4 R^2$, at higher-order in $\Omega$ than the in-plane linker shear cost.}.  It is clear from this geometry (Fig. \ref{fig: model}) that $\Delta_\perp$
is determined by the {\it relative rotation} of the grooves with respect to the
bundle lattice directions.  This relative rotation can be measured by the angle,
\begin{equation}\label{eq:psi}
  \psi(z) = \phi(z) - \Omega z 
\end{equation}
which gives the angle between the helical groove and the nearest neighbor
lattice separation as shown in Fig. \ref{fig: model}.  From this we have the
in-plane linker shear,
\begin{equation}
  \label{eq: dperp}
  \Delta_\perp = 2 a \sin \psi(z) \simeq  2 a \psi(z) .
\end{equation}
When both $\phi'$ and $\Omega$ are non-zero, the distance between crossover
points is determined by the distance over which the {\it relative angle} rotates
by $60^\circ$.  We will call this distance, $\ell$, and refer to such a
$60^\circ$ wedge shared by 2 filaments as a ``binding zone'' (see Fig.~\ref{fig:
  bindingzone}).  It is straightforward to see that,
\begin{equation}
  \ell = \frac{ \pi}{ 3 \langle \psi' \rangle} ,
\end{equation}
where $\langle \psi' \rangle$ is the mean rate of relative rotation in this
span.  Likewise, it is easy to see that of this zone, only a length $\rho \ell$
centered around the crossover point will be occupied with crosslinks.

Before analyzing the structural and thermodynamic consequences of in-plane shear
costs of crosslinking bonds, we note that a superhelical twist of the bundle
introduces other, out-of-plane modes of crosslinker shear.  We discuss, in the
Appendix, that a reorganization of the crosslinking sites along the long axis of
the bundle allows linkers to trade a high elastic energy out-of-plane shear cost
for lower energy in-plane distortion.  Furthermore, under this linker
reorganization the elastic cost of non-zero out-of-plane shear, ultimately
contributes to the total free energy of the bundle at higher order than the
leading-order, $(\psi')^2$, cost of in-plane shear derived below and represents
nominal modification to the forgoing analysis.

\section{Elastic Free Energy of Helical Filament Bundle Twist}\label{sec:elastic-free-energy}

In this section, we consider a helical filament bundle at a fixed, constant rate
of bundle twist, $\Omega$. The aim is to integrate out the distribution of the individual
crosslinks and the twist of filaments within a binding zone to derive an effective free energy in terms of bundle twist alone.
>From this, we learn that crosslinking helical filament bundles necessarily
introduces intrinsic torques which tend to wind the bundle superhelically in
order to reduce in-plane shearing of linkers.

\begin{figure}
  \center \includegraphics[width = 2.5in]{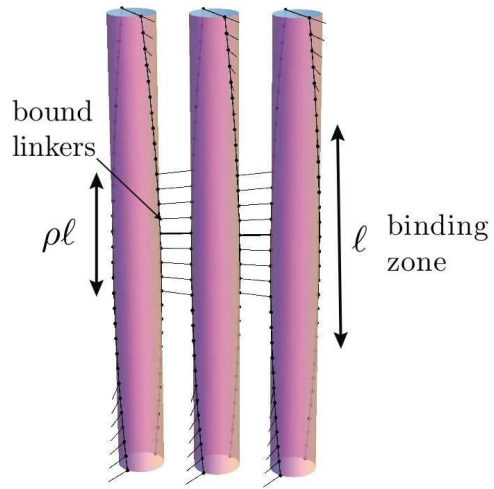}\caption{The binding
    zone is defined as the length $\ell$, over which the helical groove rotates
    by the angle $\pi/6$, which is the angle between two lattice directions in
    the hexagonal lattice. Crosslink sites are perfectly aligned, when the
    grooves of two neighboring filaments point towards each other, shown as a
    bold link in the figure. For a given linker density $\rho$, the optimal
    binding configuration is obtained, when crosslinks ``fill-up'' the binding
    zone starting from these aligned sites.  }
  \label{fig: bindingzone}
\end{figure}

Competing with the cost of crosslink shear, is the cost for twisting individual
filaments away from their native helical symmetry,
\begin{equation}
  \label{eq: torsion}
  E_{torsion} = \frac{C}{2} \int dz (\phi' - \omega_0)^2 = \frac{C}{2} \int dz \Big[ |\psi'|^2 - 2 \psi' (\omega_0 -\Omega) + (\omega_0 - \Omega)^2 \Big].
\end{equation}
Here $C$ is the torsional modulus of the filament and we used eq.~(\ref{eq:psi})
to rewrite the filament rotation, $\phi$, in terms of the groove alignment angle
$\psi$.  We analyze the respective costs of linker shear and filament twist, by
considering the profile of crosslinking occurring within a single binding zone
(shown in Figure \ref{fig: bindingzone}).  Within a binding zone $\psi(z)$
rotates by $60^\circ$ as opposing grooves come into near registry.  Thus, if
$\ell$ is the length of the binding zone along the long axis of the bundle, then
the cross term in eq.  (\ref{eq: torsion}) proportional to
$\int_{-\ell/2}^{\ell/2} dz ~ \psi' = \psi(+\ell/2)-\psi(-\ell/2)= \pi/3$ is
fixed.

As shown in Fig. \ref{fig: bindingzone}, at fixed fraction of bond crosslink
sites, $\rho$, linkers occupy a span of size $\rho \ell$ centered around the
close contact between opposing helical grooves.  Hence, for $\ell$ fixed, we may
write the $\psi$-dependence of the elastic energy of the binding zone (per
filament) in terms of the following functional,
\begin{equation}
  F (\ell) =\frac{1}{2} \int_{-\ell/2}^{\ell/2} dz \Big[ C |\psi'|^2 + \Gamma |\psi|^2 \theta (\rho \ell/2 -|z| ) \Big] .
\end{equation}
where from eq. (\ref{eq: dperp}) we find that
\begin{equation}
  \Gamma = 4 k_\perp a^2 \sigma,
\end{equation}
represents the effective pinning strength of the linkers located at $|z| \leq
\rho \ell/2$, which favor pinning the relative orientation to $\psi =0$.  The
minimal energy torsional state of the filament is determined by,
\begin{equation}
  C \frac{ \partial^2 \psi}{ \partial z^2} = \left\{ \begin{array}{ll} \Gamma \psi , & |z| \leq \rho \ell/2 \\ 0 , & |z| > \rho \ell/2 \end{array} \right.
\end{equation}
This equation, along with the boundary condition $\psi(\pm \ell/2) = \pm \pi/6$,
is satisfied by the following rotation profile,
\begin{equation}
  \psi(z) = \left\{ \begin{array}{ll} \psi_0 \sinh (\frac{ z}{\lambda}) , &  |z| \leq \rho \ell/2 \\  \psi_0 \cosh (\frac{ \rho \ell }{2 \lambda} ) \frac{(z- \rho \ell/2)}{\lambda} + \psi_0 \sinh (\frac{ \rho \ell }{2 \lambda}), & |z| > \rho \ell/2 \end{array} \right. ,
\end{equation}
where
\begin{equation}
  \psi_0 = \frac{ \pi /6}{ \cosh( \frac{ \rho \ell}{2 \lambda}) \frac{ (1-\rho) \ell}{2 \lambda} + \sinh ( \frac{ \rho \ell }{2 \lambda}) } .
\end{equation}
Here, $\lambda = \sqrt{C/\Gamma}$ is a characteristic lengthscale defined by the
relative cost of filament twist and linker shear.  It is the lengthscale
  over which the filament twist can adjust from its native rate to the value
  required in the binding zone.

\begin{figure*}
  \center \includegraphics[width = 6in]{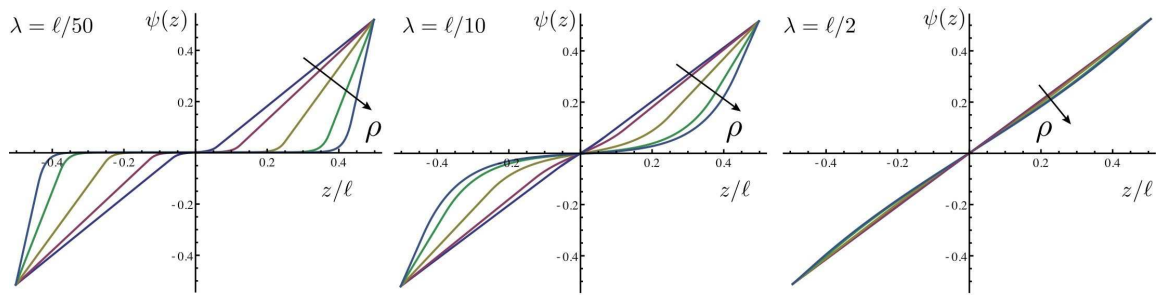}\caption{Plots of groove-slip
    profile for three different groove-slip lengths $\lambda$ and a range of
    bound linker fractions: $\rho = 0.125, 0.25, 0.5, 0.75$ and $0.875$. (Left)
    limit of groove-locked, (center) intermediate and (right) limit of
    groove-slip.  }
\label{fig: psi}
\end{figure*}

The twist profile within a binding zone is largely determined by the ratio of
this elastic lengthscale to the size of the binding zone.  Some characteristic
groove-slip profiles $\psi(z)$ are displayed in Figure \ref{fig: psi}.  In the
rigid filament limit, where $\ell /\lambda \ll 1$, the solution is simply
  $\psi (z) \simeq (\pi z / 3 \ell)$, which is equivalent to a homogeneously
  twisted state, $\phi' = \psi'+\Omega=\rm const$. This indicates a weak
modification of the filament twist due to crosslinker elasticity and we refer to
this as the limit of {\it groove-slip}.

In the opposite, rigid-crosslink limit, when $\lambda \ll \rho \ell$, we can
show that
\begin{equation}
  \psi(z) \simeq \left \{ \begin{array}{ll} 0, &  |z| \leq \rho \ell/2 \\
      \frac{\pi}{3} \frac{ z - \rho \ell /2 }{ \ell (1- \rho)} , &   |z| > \rho
      \ell/2 \end{array} \right.  
\end{equation}
In this second situation, referred to as the {\it groove-locked} limit, the
linker elasticity pins or locks the groove over the region of occupied linkers.
As a consequence the filament twist in this region follows the bundle
  twist, $\phi'=\Omega$. The rotation of the groove towards the next binding
  partner is achieved over the reduced distance $(1-\rho) \ell$, where there are
  no crosslinks.  There, the filaments rotate freely according to a 
  constant twist, $\phi' = \langle \psi' \rangle/(1- \rho) + \Omega$. 
 Accordingly, the filament twist energy will be minimized  when
  $\Omega=\omega_0$ and $\langle \psi' \rangle = 0$.  We find below that this twist profile results in a
  tendency to twist the bundle as a whole.

The minimal energy solution for $\psi(z)$, yields the following elastic energy for fixed $\ell$,
\begin{equation}
  F(\ell)/\ell = \frac{\Gamma}{2} \Big(\frac{\pi}{6}\Big)^2\frac{ \sinh(\rho
    \bar{\ell})/\bar{\ell} + (1-\rho) \cosh^2(\frac{ \rho \bar{\ell} }{2} )}{
    \Big[ \cosh (\frac{ \rho \bar{\ell}}{2} )\bar{\ell} (1-\rho)/2 +
    \sinh(\frac{ \rho \bar{\ell}}{2} ) \Big]^2} , 
\end{equation}
where $\bar{\ell} = \ell / \lambda$.  Recalling that the mean rate of twist is
related to the binding zone size by, $\langle \psi' \rangle = 2 \pi / (6 \ell)$,
the $F(\ell)/\ell $ has the following limits for groove-slip,
\begin{equation}
  \lim_{\rho \bar{\ell} \ll 1} F(\ell)/\ell =  \frac{C}{2} \langle \psi '
  \rangle ^2 +  \frac{\Gamma}{2} \Big(\frac{\pi}{6}\Big)^2  \frac{ \rho^3}{3} ,  
\end{equation}
and for groove-locked,
\begin{equation}
  \label{eq: groovelock}
  \lim_{\rho \bar{\ell} \gg 1} F(\ell)/\ell =  \frac{C}{2}\frac{ \langle \psi ' \rangle ^2}{ (1-\rho)} , 
\end{equation}
The physical interpretation underlying each limit is straightforward.  In the
groove-slip regime the elastic energy is computed from the homogeneous rotation
profile, $\psi(z) = \langle \psi ' \rangle z$ which incurs a torsional cost per
unit length $\frac{C}{2} \langle \psi ' \rangle ^2$, as well as the cost of
shearing the bound crosslinks according to $ \Gamma \int_{-\rho \ell/2}^{\rho
  \ell/2} dz ~ |\langle \psi ' \rangle z|^2 \propto \langle \psi ' \rangle^2
(\rho \ell)^3 \propto \rho^3 \ell$.  In the groove-locked regime, $\psi =0$
where crosslinks are bound and the $\pi /3$ rotation is carried over the unbound
length, $(1-\rho) \ell$, of the binding zone for which $\psi' = \langle \psi '
\rangle /(1-\rho)$.  Hence, the mean torsional energy of eq. (\ref{eq:
  groovelock}).

\begin{figure}
  \center \includegraphics[width =
  3.5in]{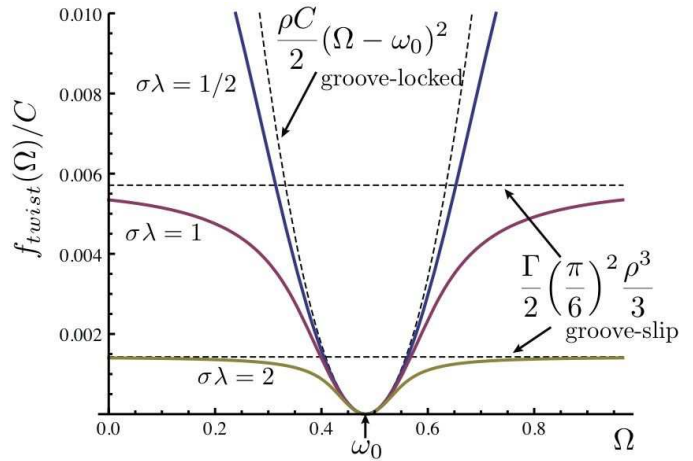}\caption{Coarse-grained elastic free energy
    $f_{twist}(\Omega)$ as determined from eq. ({eq: fgroove}). The limiting forms of
    groove-locked and groove-slip are indicated as dashed lines.}
\label{fig:torque}
\end{figure}

The final step in minimizing the elastic energy over the filament torsion is
accomplished by minimizing the combined elastic and torsional energy density
over $\ell$ for fixed $\Omega$ to find the dependence of linker-mediated groove
interactions on bundle twist,
\begin{equation}
\label{eq: fgroove}
f_{twist} (\Omega) = {\rm min}_{\ell} \Big [F(\ell)/\ell - C \langle \psi' \rangle (\omega_0 - \Omega) +\frac{C}{2}  (\omega_0 - \Omega)^2 \Big].
\end{equation}
In the limits described above it is is straightforward to show,
\begin{equation}
\langle \psi' \rangle_* \simeq \left \{ \begin{array}{ll} \omega_0 - \Omega , & \rho \ell_* \ll \lambda \\  (\omega_0 - \Omega)(1-\rho), & \rho \ell_* \gg \lambda \end{array} \right. 
\end{equation}
where again, $\ell_* = \pi/3 \langle \psi' \rangle_*$.  The
second-line above shows the clear preference of the groove-locked regime to
maintain contact between crosslinked and nearly parallel grooves on neighbor
filaments over large distances, $\ell_* \to \infty$, as
$\rho \to 1$.  Inserting these limiting cases into eq. (\ref{eq: fgroove}) we find a central result of our analysis, the free energy of linker and filament elasticity in terms of $\Omega$,
\begin{equation}
\label{eq:ftwist}
f_{twist} (\Omega) \simeq \left \{ \begin{array}{ll} \frac{\Gamma}{2}
    (\frac{\pi}{6})^2  \frac{ \rho^3}{3} , & \rho \ell_* \ll \lambda \\ 
    \rho    \frac{C}{2} (\Omega - \omega_0)^2, & \rho \ell_* \gg \lambda \end{array}
\right.  
\end{equation}
The full non-linear behavior for
the coarse-grained free energy (as calculated from eq.  (\ref{eq: fgroove})) is
shown in Figure~\ref{fig:torque}. In the groove-slip limit ($\rho \ell_* \ll
\lambda $) the linker-mediated groove interactions become insensitive to bundle
twist, as the crosslinks are not stiff enough to noticeably affect the state
  of twist of the individual filaments. In this case, the elastic energy is set by the
  deformation of the flexible crosslinks, $f_{twist}\sim \Gamma$, but does not
  depend on the amplitude of the bundle twist $\Omega$.  Thus, increasing bundle
  twist does not appreciably increase the cost of shear deformation in the crosslinks.
  Instead, the crosslinks reorganize into new binding sites allowing them to maintain
  a constant average crosslink deformation.  This is possible, as in hexagonal
  bundles the angle $\psi = 2\pi/3$ is an upper limit for the angle between two
  neighboring grooves, {\it independent} of the size of the binding zone, $\ell$. Accordingly, the shear deformation $\Delta_\perp =
  2a\psi$ of the maximally stretched linker cannot grow larger than $2 a \rho \pi/3$, independent of the bundle twist
  $\Omega$.

  In the groove-locked limit ($\rho \ell_* \gg \lambda $) the energy scale is
  set by the filament twist stiffness $C$. The linker-mediated interactions lead
  to a preference to twist the bundle at a rate equal to the intrinsic twist of
  filaments, $\Omega\sim\omega_0$.  This latter case reflects the fact that
  crosslinks on neighboring filaments lock a span of length $\rho \ell$ into a
  parallel configuration.  By rotating the interfilament position at a rate
  $\Omega = \omega_0$, these groove-locked domains on neighbor filaments can be
  brought into coincidence with the native, helical geometry of the untwisted
  grooves.  Therefore, a high degree of crosslinking by rigid linkers ($\rho
  \ell_* \gg 1$) induces an {\it intrinsic torque} on the entire bundle, which
  prefers the fiament lattice to rotate at the rate of the helical grooves.  The
  crossover between groove-locked and groove-slip behavior can be related to
  $\delta \Omega = \Omega - \omega_0$.  Groove-locking occurs for $|\delta
  \Omega| \lesssim \delta \Omega_c$, while groove-slip occurs for larger
  deviations from optimal twist, $|\delta \Omega|$, where,
\begin{equation}
\label{eq: domc}
\delta \Omega_c \equiv \rho \frac{  \pi }{6\sqrt{3} \lambda} .
\end{equation} 
Therefore, as shown in Fig. \ref{fig:torque}, the harmonic,
linear-elastic twist dependence of the linker elastic energy is maintained for
bundle twists near to the intrinsic rate of groove twist, $\omega_0$, over range
of twists that increases with the fraction of bound crosslinks.

\section{Properties of Bundles with Crosslink Induced Torque}\label{sec:prop-bundl-with}

In the previous section we have derived the form of the linker-induced elastic
energy for global twist of crosslinked bundles of helical filaments.  We found that crosslinks induce torques that tend to twist the bundle as a whole. These intrinsic torques give rise to unusual structural and mechanical
properties for self-assembled bundles, including a highly non-linear dependence
of bundle twist on lateral size and externally applied torsional moment. 

To demonstrate these properties, we consider crosslinked bundles of
semi-flexible helical filaments with a twist-dependence described by
eq.~(\ref{eq: fgroove}).  The additional mechanical costs of filament bending
can be described by the following simple elastic energy,
\begin{equation}
  E_{bend} = \frac{K}{2} \sum_i \int ds ~ \kappa^2(\rv_i) . 
\end{equation}
where $K$ is the bend modulus and $\kappa \simeq \Omega^2 |\rv_i|$ is the
curvature of the $i$th filament in a twisted bundle.  Averaging over a
cylindrical cross-section of radius $R$, we obtain a free energy per unit
filament length,
\begin{equation}\label{eq:fomega}
  f (\Omega) = f_{twist} (\Omega)+ \frac{K}{4} \Omega^4 R^2 .
\end{equation}

Analytical progress is hampered by the complicated form of the twist free
energy, Eq.~(\ref{eq: fgroove}). In the following we therefore assume the
simplified expression
\begin{equation}
\label{eq: interp}
  f_{twist} (\Omega) =\frac{ \rho C \delta \Omega_c^2}{2}\Big[1-e^{-(\Omega - \omega_0)^2/\delta \Omega_c^2} \Big] ,  
\end{equation}
which captures the limiting groove-locked and groove-slip behaviors of
eq.~(\ref{eq:ftwist}) as well as the overall non-linear dependence on $\Omega$
between these two limits.

In the following subsection, we analyze the structural and mechanical properties
of bundles in terms of two dimensionless parameters, $\bar{R} \equiv R/R_{un}$
and $\delta \bar{\Omega}_c \equiv \delta \Omega_c / \omega_0$, where
\begin{equation}
  R_{un} = \sqrt{\frac{\rho C}{ K \omega_0^2}},
\end{equation}
is a characteristic bundle size at which the mechanical cost of filament bending
becomes comparable to the linker-induced cost of twisting the filaments from their native symmetry.

\begin{figure}
  \center \includegraphics[width = 3in]{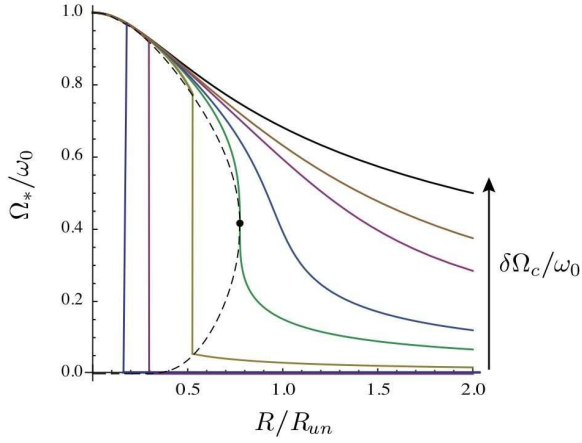}\caption{Plots of the
    equilibrium degree of bundle twist, $\Omega_*$, vs. bundle twist, $R$, for
    $\delta \Omega_c / \omega_0 = 0.1, 0.2, 0.3, \delta \bar{\Omega}_c \simeq
    0.362, 0.4, 0.5, 0.6$ and $\infty$, the groove-locked limit described by eq. (\ref{eq: Omlocked}).  The filled
    circle shows a critical point at $\delta \bar{\Omega}_c \simeq 0.362$,
    $\bar{\Omega}_* \simeq 0.407$ and $\bar{R} = 1.073$, while the dashed line
    shows region of first-order jumps in $\Omega_*$ for $\delta \bar{\Omega}_c <
    0.362$.  }
\label{fig: OmvsR}
\end{figure}

\subsection{Size dependent bundle twist}

We first analyze the equilibrium twist of self-assembled bundles as function of
lateral radius, $R$.  According to $f_{twist} (\Omega)$ crosslink-mediated
torques prefer a constant degree of twist, $\Omega = \omega_0$. On the other
hand, bending resistance of filaments penalizes bundle twist in an $R$-dependent
manner due to the linear increase of filament curvature with radial distance
from the helical bundle center.

Based on our expression for $f_{twist}(\Omega)$, For a fixed $R$ and $\rho$, the
equilibrium bundle twist is determined from the solutions to
\begin{equation}
  (\rho C)^{-1} \frac{\partial f}{ \partial \Omega} =  (\bar{\Omega} - 1)
  e^{-(\bar\Omega - 1)^2/\delta \bar{\Omega}_c^2} + \bar{\Omega}^3 \bar{R}^2
  = 0 . 
\end{equation}
where $\bar{\Omega} \equiv \Omega/\omega_0$ is the reduced twist.  For the
limiting case of $\delta \bar{\Omega}_c \gg 1$, where linker shear is
sufficiently strong to maintain the high torque, groove-locked behavior over the
entire range $0 \leq \bar{\Omega} \leq 1$, the equilibrium bundle twist
satisfies a cubic equation,
\begin{equation}
  (\bar{\Omega} - 1) + \bar{\Omega}^3 \bar{R}^2 = 0 ,\  {\rm for} \ \delta \bar {\Omega}_c \gg 1 .
\end{equation}
In this limit, the $R$ dependence of the twist has the form
\begin{equation}
\label{eq: Omlocked}
\bar{\Omega}(\bar{R}) = \frac{1}{\sqrt{3} \bar{R} } \Big[ x^{1/3}(\Bar{R}) -
x^{-1/3}(\Bar{R}) \Big] ,\  {\rm for} \ \delta \bar {\Omega}_c \gg 1 ,
\end{equation}
where
\begin{equation}
  x(\bar{R}) = \sqrt{\frac{27}{4} } \bar{R} + \sqrt{1+ \frac{27  \bar{R}^2 }{4} } .
\end{equation}
In this groove-locked limit, the bundle is unwound continuously as $R$ increases
due to the increased cost of filament bending.  In the limit of large bundles,
$\bar{R} \gg 1$, eq. (\ref{eq: Omlocked}) predicts a power law dependence of optimal
twist on bundle size, $\bar{\Omega}_* \sim \bar{R}^{-2/3}$. 

The range of groove-locked behavior is highly sensitive to the degree of
crosslinking as $\delta \Omega_c \propto \rho$, hence for weakly crosslinked
bundles as $\rho \to 0$ the torsional energy dependence necessarily crosses over
to groove-slip behavior, becoming largely insensitive to $\Omega$.  Thus, in the
most general case, we expect the groove-locked predictions of $\Omega_*$
described by eq. (\ref{eq: Omlocked}) to hold only for sufficiently small
bundles where the groove-locked approximation predicts that $1 -
\bar{\Omega}_*(\bar{R}) \lesssim \delta \bar{\Omega}_c$.  Beyond this size, we
expect the decrease of $\Omega_*$ with increasing size $R$ to become more rapid
as the bundle slips from the high-torque region.

The full dependence of $\Omega_*$ vs. $R$ is shown in Fig. \ref{fig: OmvsR} for
several values of $\delta \bar{\Omega}_c < 1$.  These show that unwinding of the
bundle due to the increased bending cost of large bundles becomes more rapid in
comparison to the predictions of eq.
(\ref{eq: Omlocked}) as $\delta \bar{\Omega}_c$ is decreased.  For a critical
value, $\delta \bar{\Omega}_c = \sqrt{\frac{11 \sqrt{33}}{32} - \frac{59}{32} }
\simeq 0.362$, the sensitivity of twist to bundle size becomes singular,
$\frac{\delta \Omega_*}{ \delta R} \to - \infty$, at $\bar{R} \simeq 1.0731$.
For $\delta \bar{\Omega}_c$ below this critical value, $\Omega_*$ unwinds by a
discontinuous, first-order jump, as the elastic energy minimum slips from the
narrow high-torque behavior near $\Omega \approx \omega_0$ to the low-torque,
groove-slip behavior of $f_{twist} (\Omega)$ for $\omega_0 - \Omega \lesssim
\delta \Omega_c$.

In this regime, the equililbrium state of twist is therefore highly susceptible
to small changes in the mechanical or geometrical properties of the bundle.
Changing only slightly bundle radius or crosslink density may result in a sudden
and strong reduction of the overall bundle twist $\bar\Omega$.

\begin{figure*}
  \begin{center}
    \includegraphics[width=0.7\columnwidth,angle=0]{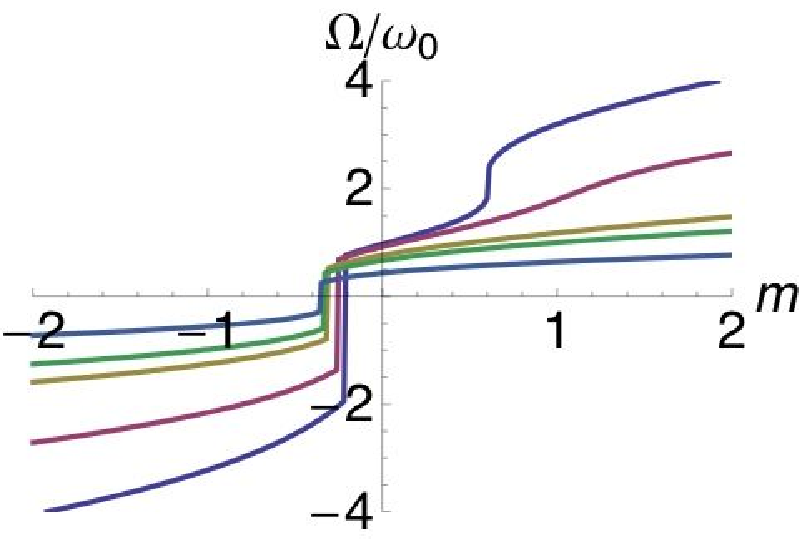}
    \includegraphics[width=0.7\columnwidth,angle=0]{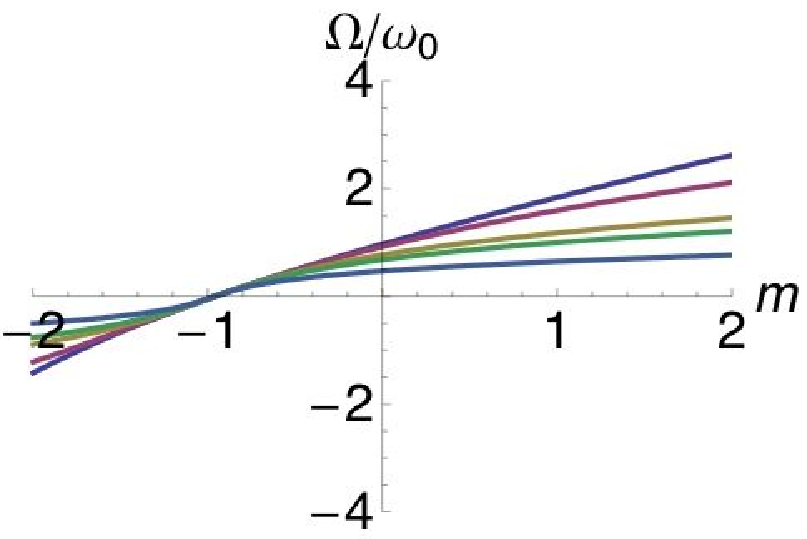}
  \end{center}
  \caption{Bundle twist $\Omega$ as a function of reduced torque $m=M/\rho
    C\omega_0$, for different $\bar R^2=0.03,0.1,0.5,1.0,5.0$. (left) A small value
    of $\delta\bar\Omega_c=1$ allows to observe the discontinuous transition
    from groove-locked to slip as the external torque is increased. (right) For
    a larger value of $\delta\bar\Omega_c=5$ the torque-twist relation is
    completely smooth.}\label{fig: externaltorque}
\end{figure*}

\subsection{Crosslinked bundles of helical filaments subject to external torque
}

In this section, we demonstrate how this sensitivity leads to a highly
non-linear twist response when the bundle is subject to an external torque.
Experiments that probe the torque-twist relation of single molecules have proven
fruitful tools in advancing the understanding of the mechanical properties of
biopolymers, such as DNA~\cite{nelson}. While the active manipulation of
filament bundles is still a very delicate task \cite{strehle}, experiments in
this direction may soon be feasible and may therefore provide an indirect means
to probe the elastic properties of crosslinkers and their interactions with
filaments.

In the presence of an external torque $M$, the free energy Eq.~(\ref{eq:fomega})
becomes
\begin{eqnarray}\label{eq:ftorque}
  f(\Omega) = f_{\rm twist}(\Omega) + \frac{K}{4}\Omega^4R^2-M\Omega\,.
\end{eqnarray}

In the absence of the filament bending term $\sim K\Omega^4$, it is
straighforward to see that the bundle is thermodynamically unstable for any
nonzero value of $M$. The linear term $-M\Omega$ leads to a tilting of the twist
free energy $f_{\rm twist}(\Omega)$.  As $f_{\rm twist}$ is asymptotically
independent of $\Omega$, in the absence of bending resistance, the thermodynamic
groundstate is always the fully twisted state, $\Omega\to\infty$. Physically,
this means that increasing the external torque does not lead to restoring
forces, e.g. in the form of crosslink shearing.  Instead, the bundle adapts to
the increased load by a reorganization of the crosslinks into new binding sites.

Accounting for the mechanical cost of filament bending, the bundle is stabilized
at a value of $\Omega$, at which the twist-induced bending energy of the
filaments balance the external torque. The full dependence of $\Omega$ vs. $M$
is shown in Fig.~\ref{fig: externaltorque}.  For small values of
$\delta\bar\Omega_c$ and $\bar R$ the instability appears as a sudden change of
$\Omega$ with the external torque $M$.  Note that due to the helical nature of
the filaments, the response of the bundle will be inherently asymmetric and
different for positive or negative torques.  Hence, careful measurements of the
non-linear torsional response of the self-assembled bundles should provide an
indirect, experimental means to probe the twist state of crosslinked filaments.

\section{Linker-mediated filament assembly}\label{sec:link-medi-filam}

\begin{figure*}
  \begin{center}
    \includegraphics[width=0.65\columnwidth,angle=0]{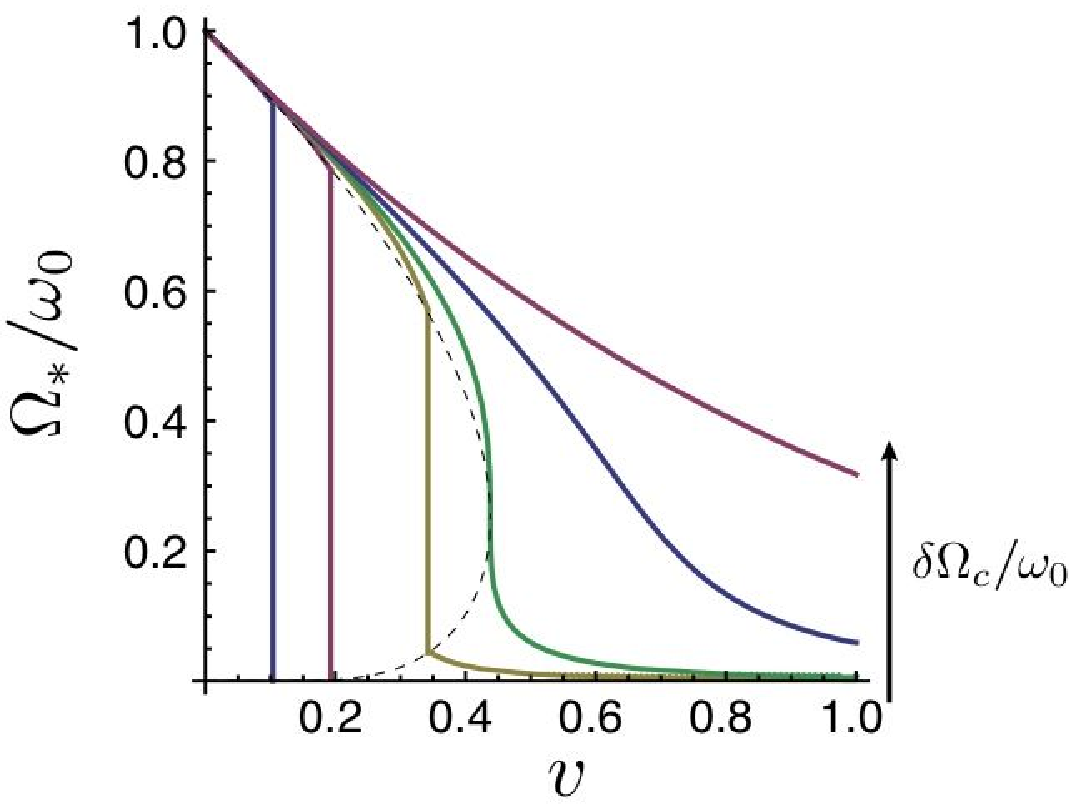}
    \includegraphics[width=0.7\columnwidth,angle=0]{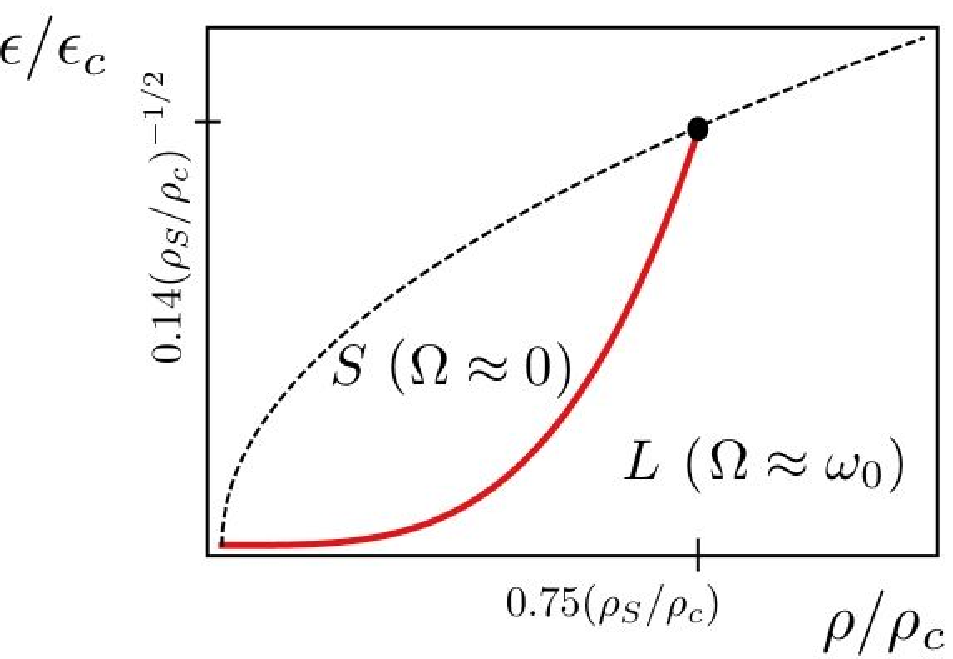}
  \end{center}
  \caption{(left) Bundle twist $\Omega$ in thermodynamic equilibrium as function
    of $v$ for various value of $\rho_S/\rho=0,1,1.333,1.5,2,3$. (right) A
    sketch of the resulting phase diagram indicating the discontinuous
    transition from groove-locked (L) to groove-slip (S).  The critical point
    shifts along the dashed line, when the effective stiffness of crosslinks, as
    characterized by $\rho_S/\rho_c$ is changed.  }\label{fig.omega.rhoS}
\end{figure*}

In the previous sections we have discussed the properties of bundles of a given
radius $R$.  In this section we analyze the equilibrium thermodynamics of a
system of self-assembled filaments and consider the thermodynamic stability of
bundles of finite radius in the presence of a fixed degree of crosslinking.

Here, we consider a system possessing a fixed total number of filaments.  In the
bundled state, all filaments are assumed to form bundles of a mean-size, $R$,
and a negligible number of unbundled filaments remain dispersed in solution.  As
described in eq. (\ref{eq: elink}), each bound crosslink contributes,
$-\epsilon$, of cohesive free energy.  In the bulk of a self-assembled
aggregate, crosslinks contribute $-3 \rho \epsilon \sigma$ per unit filament
length, as there are 3 crosslinking ``channels'' per filament in the interior of
the bundle.  At the outer boundary of the bundle, there are fewer crosslinks,
roughly $3/2 \rho \sigma L$ fewer cohesive bonds times the number of filaments
at the surface of the bundle, $N_s$.  For a large bundle with a circular
cross-section, we may estimate $N_s \simeq 2 \pi R/d$ so that the net cohesive
free energy per unit filament length from crosslinked bundle of size $R$ has the
form,
\begin{equation}
  F_{\rm coh}/N_fL= -3\rho\epsilon\sigma + \frac{ 3 \rho \sigma \epsilon}{\rho_0 d R},
\end{equation}
where $\rho_0= d^{-2}/\sqrt{3}$.  Combining the cohesive energy with the elastic
costs of linker shear and filament bending we may write the total free energy
per unit filament length,
\begin{eqnarray}\label{eq:f.all.terms}
  f(\Omega,R) = f_{\rm 
    twist}(\Omega)+\frac{\rho C \omega_0^2}{2} \Big[ \frac{\Omega^4 R^2
    (\rho_c/\rho) }{2 \omega_0^4 d^2} + \frac{(\epsilon/\epsilon_c) }{ d  \rho_0
    R} - \epsilon/\epsilon_c \Big], 
\end{eqnarray}
where the ratio of bend to twist moduli define a characteristic crosslink
fraction,
\begin{equation}
  \rho_c = \frac{K (\omega_0 d)^2}{C},
\end{equation}
and
\begin{equation}
  \epsilon_c = \frac{  C \omega_0^2 }{6 \sigma},
\end{equation}
is a characteristic cohesive energy scale.  The thermodynamics of filament
assembly are characterized by the dependence of the equilibrium, free energy
minimizing values of $R$ and $\Omega$ on $\rho$ and $\epsilon$.  It is
straightforward to show that the value of $R$ minimizing $f(\Omega,R)$ at fixed
$\Omega$ is determined by a balance between the cohesive and bending energies of
the bundle, satisfying
\begin{equation}
  \label{eq: RofOM}
  R_*^3 (\Omega)= \frac{ d}{\rho_0 (\Omega/\omega_0)^4} (\epsilon/\epsilon_c) (\rho/\rho_c) .
\end{equation} 
Thus, the growth of equilibrium radius is quite generally correlated to
decreasing twist for self-assembled bundles, and a non-zero measure of bundle
twist implies a finite equilibrium radius.  

Using the definition of $\delta\Omega_c$, Eq.(\ref{eq: domc}), we can define
another characteristic crosslink fraction,
\begin{equation}
  \rho_S \equiv \frac{ \omega_0 \rho}{ \delta \Omega_ c}= \frac{ 6 \sqrt{3}}{\pi}\omega_0 \lambda 
\end{equation}
that characterizes the crosslink density below which the bundle ``slips'' to the
low-torque elastic energy at $\Omega = 0$. With the solution for $R_*$,
Eq.~(\ref{eq: RofOM}), and the form of $f_{twist}$ from eq. (\ref{eq: interp}),
we may rewrite the reduced energy density purely in terms of the reduced twist,
$\bar{\Omega}$,
\begin{equation}
\label{eq: reduced}
\frac{ f(\bar{\Omega})}{ \rho C \omega_0^2} = \frac{(\rho/\rho_S)^2}{2} \Big[1- e^{-(\bar{\Omega}-1)^2/(\rho/\rho_s)^2}\Big] +\frac{3}{4} v(\rho,\epsilon) \bar{\Omega}^{4/3} - \epsilon/\epsilon_c ,
\end{equation}
where,
\begin{equation}
v(\rho,\epsilon) \equiv 3^{1/3} \frac{ (\epsilon/\epsilon_c)^{2/3} }{ (\rho/\rho_c)^{1/3} }.
\end{equation}
The reduced form of the free energy density, eq. (\ref{eq: reduced}),
demonstrates that the phase behavior of helical crosslinked bundles is
determined by two dimensionless parameters: $\rho/\rho_S$ and $v$.  Consequently
the thermodynamic dependence of the assembly on crosslink density and cohesive
energy of crosslinks is encoded in the $\rho$ and $\epsilon$ dependence of these
parameters.

The first term in eq. (\ref{eq: reduced}), representing the crosslink-mediated
torsional energy, is minimized at $\bar{\Omega} = 1$, while the second term,
representing the combined cohesive and bending energies is minimized at zero
twist, $\bar{\Omega} = 0$.  Hence, generically $f(\bar{\Omega})$ may be
characterized by two minima, whose relative depth is determined by $\rho/\rho_S$
and $v$.  It is straightforward to analyze the case of rigid linkers, for which
we expect $\rho /\rho_S \gg 1$ over the range of filament assembly and where the
first term in eq. (\ref{eq: reduced}) adopts the groove-locked limit,
$(\bar{\Omega}-1)^2/2$.  In this case, we minimize the reduced free energy
density for the limiting cases,
\begin{equation}
  \bar{\Omega}_* \simeq \begin{cases}1 - v, & {\rm for} \ v \ll 1 \ {\rm and } \ \rho \gg \rho_S \\ v^{-3} , & {\rm for} \ v \gg 1 \ {\rm and } \ \rho\gg \rho_S 
  \end{cases}
\end{equation}
we note that in the limit of $v \ll 1$ and $\rho/\rho_S \gg 1$ the minimum of
$f(\bar{\Omega})$ corresponds to the groove-locked state, $\bar{\Omega}_* \to 1$
from which we find the following dependence of equilibrium bundle size on $\rho$
and $\epsilon$,
\begin{equation}
  \label{eq: size}
  R_* \simeq d \begin{cases} (\epsilon/\epsilon_c)^{1/3} (\rho/\rho_c)^{1/3} ,  & {\rm for} \ v \ll 1 \ {\rm and } \ \rho\gg \rho_S \\ 3 (\epsilon/\epsilon_c)^{3} (\rho/\rho_c)^{-1}, & {\rm for} \ v \gg 1 \ {\rm and } \ \rho\gg \rho_S \end{cases} ,
\end{equation}
where we have used $(d/\rho_0)^{1/3} \approx d$.  Stable, microscopic bundles are associated with
the limit of high cross-link density and relatively weak cohesive energy per
link where $v \ll1$ while $\rho \gg \rho_S$.  As one might have expected, eq.
(\ref{eq: size}) suggests that due to the adhesive effect of crosslinking,
bundles grow with increasing $\epsilon$, as larger bundles imply smaller surface
effects.  Perhaps more surprising, the $v \ll 1$ limit of this model predicts
that at large linker densities bundles assemble to a finite size that {\it
  grows} with increasing fraction of bound crosslinkers, growing as $R_* \sim
\rho^{1/3}$.  In this regime the bundle is fully twisted, so filaments have
  to bend in order to be incorporated into the bundle. Increasing bundle size is
  therefore only possible when the mechanical cost of filament bending is offset by the cohesive energy gain of adding crosslinks.

In the limit of large $v$, note that the size of equilibrium bundles also
diverges in the limit of small linker fraction as $R_* \sim \rho^{-1}$,
indicating a smooth crossover to a state of unlimited, macroscopic filament
assembly in the $\rho \to 0$ limit.

It is straightforward to determine the full equation of state for an arbitrary
value of $\rho_S$, relating equilibrium twist, $\bar{\Omega}_*$, to $\rho$ and
$\epsilon$ from $d f/d \bar{\Omega} = 0$,
\begin{equation}\label{eq:eos}
  \Big(\frac{\rho}{\rho_S}\Big)^2 =- \frac{(1-\bar{\Omega}_*)^2}{\ln \big[ v
    \bar{\Omega}_*^{1/3} /(1-\bar{\Omega}_*) \big] } . 
\end{equation}
The solutions to this equation of state are shown in Fig.~\ref{fig.omega.rhoS},
where we plot $\Omega_*$ as a function of $v$ for constant values of
$\rho/\rho_S$. These results show that equilibrium twist decreases from
$\omega_0$ both with {\it decreased} $\rho/\rho_S$ as well as with {\it
  increased} $v$.  Underlying the unwinding of equilibrium bundles are two
effects, driven by decreasing crosslink fraction.  First, as the number of
crosslinkers in the bundles is reduced, the strength of the intrinsic torques
that drive the bundle towards $\bar{\Omega}_*=1$ is correspondingly reduced.
Second, $v$ diverges as $\rho \to 0$ indicating a dramatic increase in the
relative importance of filament bending in the elastic energy, further enhancing
the preference to untwist the bundle.  According to eq. (\ref{eq: RofOM})
equilibrium bundles grow in radius as they unwind, and ultimately diverge in
size at the untwisted, $\Omega = 0$ state.

\begin{figure*}
 \begin{center}
   \includegraphics[width=0.75\columnwidth,angle=0]{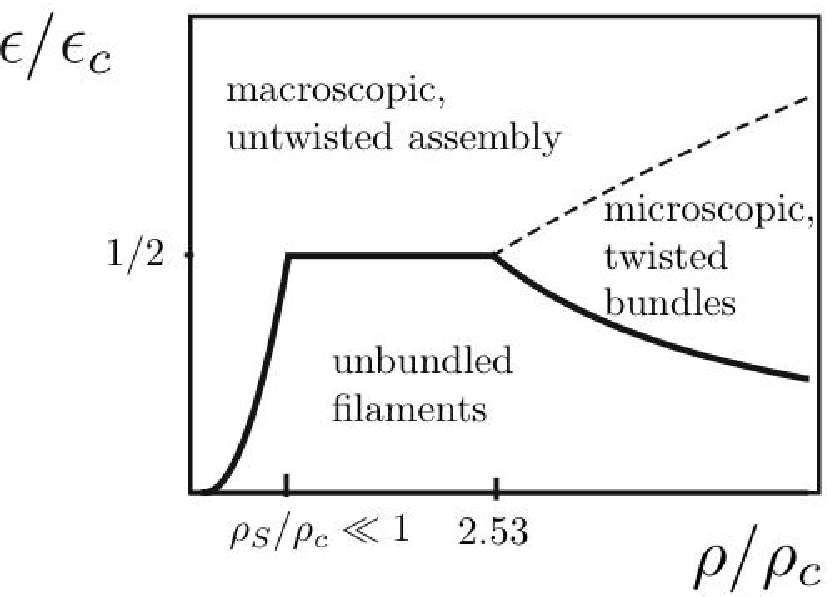}
   \includegraphics[width=0.75\columnwidth,angle=0]{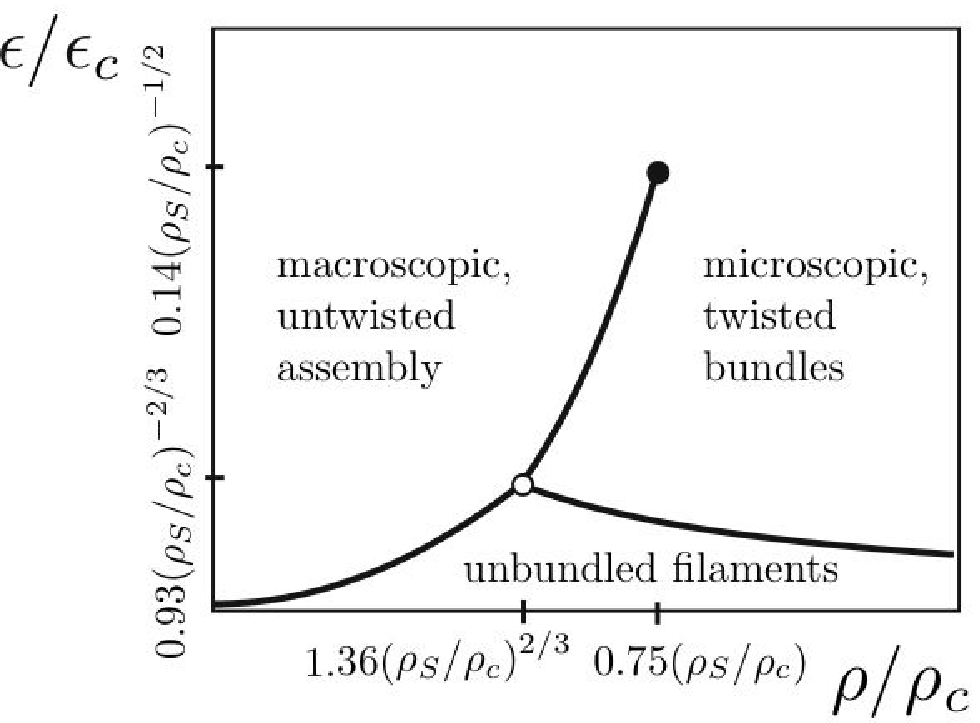}
\end{center}
\caption{Sketches of the two scenarios for the state diagram of bundle assembly
  in the $\epsilon-\rho$ plane: rigid linkers, $\rho_c/\rho_S\gg 1$ (left); and
  flexible linkers, $\rho_c/ \rho_S \ll 1$ (right).  For $\rho_S<2.53\rho_c$ the
  critical point is not relevant as it would lie deep inside the filament phase
  (left). The dashed line indicates the crossover between microscopic,
  finite-sized bundles and macroscopic bundles with diverging size. (right) For
  $\rho_S>2.53\rho_c$ the critical point $(\rho_\star,\epsilon_\star)$ is inside
  the bundle phase (shown as filled circle). The triple point
  $(\rho_T,\epsilon_T)$ signals coexistence of the filament and the two
  assembled filament phases (shown as open circle).}\label{fig.phasediagram}
\end{figure*}

Fig.~\ref{fig.omega.rhoS} shows this unwinding with decreased $\rho$ may occur
either continuously or discontinuously, accompanied by a rapid jump in
equilibrium twist.  From the equation of state, Eq.~(\ref{eq:eos}), we find a
critical value $v=v_\star=3/2^{4/3}e \simeq 0.438$ that separates smooth from
discontinuous bundle untwisting. At this critical point, $\rho_\star/\rho_S
=3/4,\Omega_\star = 1/4$~\footnote{The numerical values are derived with the
  Gaussian interpolation formula for $f_{\rm twist}$ and may differ slightly in
  the full model. }.  For $v < v_*$ the equilibrium twist state jumps from the
groove-locked minimum of $f(\bar{\Omega})$ near $\Omega = \omega_0$ to the
nearly unwound groove-slip state.  This discontinuous and highly non-linear
thermodynamic dependence of self-assembled bundles derives from the non-linear
interplay between linker shear and filament twist encoded in $f_{twist}
(\Omega)$.
  
We now proceed to sketch the possible phase diagrams of self-assembled filament
systems.  We begin by analyzing a transition between the 2 torsional states of
self-assembled filament arrays, groove-locked vs. groove-slip.  In terms of
parameters $\epsilon$ and $\rho$ we sketch the twist-state phase diagram of
Fig.\ref{fig.omega.rhoS}b. Thermodynamically stable finite-sized bundles are
associated with the groove-locked regime (L), while in the slip regime (S)
radial bundle growth is nearly unlimited in equilibrium as $\Omega_* \approx 0$.
The line of discontinuous transitions between locked and slip regime terminates
at the critical point.  The transition line can approximately be calculated by
comparing the $\Omega\to \omega_0$ solution ($f/(C \rho\omega_0^2)\to3v /4 -
\epsilon/\epsilon_c$, groove-locked) with the $\Omega\to 0$ solution ($f/(C \rho
\omega_0^2)\to (\rho/\rho_S)^2 /2 - \epsilon/\epsilon_c$, groove-slip).  Note,
that this calculation will not be accurate close to the critical point, where
$\Omega_*\approx \omega_0/4$. Equating these two limiting forms of the energy,
we find a relationship between $\epsilon$ and $\rho$, satisfied at the (LS)
boundary,
\begin{eqnarray}
  \label{eq:LS}
  \epsilon_{\rm (LS)}/\epsilon_c \simeq (2/3)^{3/2} (\rho/\rho_c)^{7/2}(\rho_c/\rho_S)^{3} \qquad \rm (LS) . 
\end{eqnarray}
Across this line of first order transitions, the equilibrium twist of bundles
slips from a nearly groove-locked state with $\Omega \approx \omega_0$ of
microscopic size to a nearly untwisted and macroscopically sized bundle.  

The location of the critical point follows from the condition $v=v_\star\approx 0.438$ which gives
\begin{equation}
  \epsilon/\epsilon_c = v_*^{3/2} (\rho/\rho_c)^{1/2}.
\end{equation}
This critical branch is sketched as a dashed line in Fig.~\ref{fig.omega.rhoS}b.
Hence, with increasing $\rho_S$, say by decreasing linker shear stiffness, the
(LS) boundary and the critical point shift to larger values of $\rho$.  Notice
that the bundle phase diagram has the familiar form of a liquid-gas transition.
Stable finite-sized bundles correspond to the ``condensed'' phase, while in the
``gas-phase'' the bundles are large and nearly untwisted.  Below a critical
value of $\epsilon$ corresponding to $v_*$, there is a well-defined first-order
transition between these states at a certain linker density. For sufficiently
strong cohesive energies the state of untwisted, macroscopic assembly evolves
continuously to the state of finite-sized bundles at high linker fraction.

In addition to the ``microscopic'' and ``macrosopic'' bundle phases, we also
consider the possibility of a state of unassembled filaments, where no bundles
form and filaments remain uncrosslinked. Neglecting entropic contributions, such
as translational entropy of linkers and filaments, the free energy density in
the filament phase vanishes, $f_{\rm fil}=0$, as free filaments are mechanically
undistorted.  Thus, the state of bundled filaments is stable with respect to
unbundled filaments where the free energy of Eq.(\ref{eq:f.all.terms}) remains
negative, indicating a net lowering of the free energy due to crosslink binding.
The phase boundary, $f(\Omega_\star,R_\star)=0$, can be determined approximately
by assuming the limiting cases of groove-locked ($\Omega = \omega_0$) and
groove-slip ($\Omega= 0$).  Equating the energy density of the groove-locked
bundles to the free filament case, we find the condition satisfied along the
twisted, bundle/filament (BF) phase boundary,
\begin{equation}\label{eq: BF}
  \epsilon_{\rm (BF)}/\epsilon_c \simeq  \frac{81}{64}
  (\rho/\rho_c)^{-1} .
\end{equation}
We may also estimate the phase boundary between untwisted, macroscopic bundle
assembly and free filament phase, denoted as (UF), by comparing the free energy
of the state with $\Omega = 0$ and $R\to \infty$ to the free filament energy in
both the groove-locked and groove-slip regimes. From eq. (\ref{eq: reduced}),
we find this boundary satisfies,
\begin{equation}
  \label{eq: UF}
  \epsilon_{\rm (UF) }/\epsilon_c \simeq \begin{cases} \frac{1}{2}
    (\rho/\rho_S)^2 & {\rm for } \ \rho \ll \rho_S \\ \frac{1}{2} &  {\rm for }
    \ \rho \gg \rho_S \end{cases} . 
\end{equation}

Based on the $\epsilon$- and $\rho$-dependence of these boundaries, we sketch
two basic scenarios for the diagram of state of bundle assembly of crosslinked
helical filaments, based on the interference of the (LS), (BF) and (UF)
boundaries.  When $\rho_S < 2.53 \rho_c$, the critical point at $v_*$ and
$\rho_*$ lies deep within the filament phase and the (LS) line is not relevant.
This is the limit of effectively {\it rigid linkers}, characterized by
$\rho_c/\rho_S \gg 1$.  We sketch the phase behavior in
Fig.~\ref{fig.phasediagram}a.  Here, the boundaries (UF) and (BF) meet at the
point $\epsilon = \epsilon_c /2$ and $\rho = 81/32 \rho_c \simeq 2.53 \rho_c$.
Below this critical value of cohesive energy of crosslinks, $\epsilon_c/2$,
bundles only form at crosslinker fractions that are larger than $1.27
\rho_c(\epsilon_c / \epsilon)$.  Bundles formed for this low-cohesive energy
regime are twisted and grow in size with increasing crosslinking density as $R_*
\sim \rho^{1/3}$.  For smaller crosslink densities, the filaments remain in the
dispersed state, allowing for a narrow region of slip-induced, macroscopic
filament assembly predicted for $\rho \lesssim \rho_S$.  For $\epsilon >
\epsilon_c/2$ filaments assemble at all crosslinker fractions, with a state of
macroscopic bundle assembly crossing over to microscopic, twisted bundles as
$\rho$ increase from 0 to values of order $\rho_c$.

The second scenario, $\rho_S\geq 2.53 \rho_c$, occurs in the limit of {\it
  flexible linkers}.  In this case the critical point lies inside the regime of
linker-mediated filament assembly, as shown in Fig.~\ref{fig.phasediagram}b.  In
addition to the line of first-order transitions separating the groove-locked and groove-slip
states of bundles, we also have a triple point, at which the two bundle phases,
macroscopic and microscopic, coexist with the state of dispersed filaments. From
the scaling expressions given for (BF) and (UF) we locate the triple point at
\begin{equation}
\rho_{\rm T} /\rho_c\simeq 1.36(\rho_S/\rho_c)^{2/3}, \  \epsilon_{\rm T}/\epsilon \simeq 0.93 (\rho_S/\rho_c)^{-2/3} .
\end{equation}
In this case, bundled phases form for all cohesive energies, $\epsilon >
\epsilon_{\rm T}$, with a line of first-order transitions that separates the
phase of macroscopic, nearly untwisted bundles at low linker densities from the
phase of small radius, twisted bundles at high linker density and terminates at
the second order critical point.  For weaker cohesive energy per crosslinker,
there are three possible states of filament assembly.  At low density of bound
linkers, $\rho < \rho_{\rm(UF)}$, we predict a phase of untwisted, macroscopic
filament assembly.  For intermediate densities of linkers, $\rho_{\rm(UF) }<
\rho < \rho_{\rm (BF)}$, we predict a state of unbound and disperse linkers.
Finally, at the highest densities of linkers, $\rho> \rho_{\rm (BF)}$, we find a
re-entrant state of twisted, filament bundles, whose finite diameter grows with
linker density as $R_* \sim \rho^{1/3}$.

\section{Conclusion}

In this study, we have analyzed a coarse-grained model for crosslinking in
ordered arrays, or bundles, of helical filaments.  A primary result of this
model is a quantitative relationship between the presence of crosslinkers in
bundles, and intrinsic torques that act to coherently twist entire bundles
superhelically around their central axis.  Such global distortion naturally
competes with the mechanical cost of bending stiff filaments, providing a
complex feedback between the torsional structure, size and thermodynamics of
bundle assembly.  Along with the total number and binding affinity of
crosslinkers, the key parameters governing the structure and assembly of
crosslinked bundles include the elastic properties of the filaments and the
linkers themselves.  Indeed, we find the ratio $\lambda=\sqrt{C/\Gamma}$ of
linker stiffness $\Gamma$ to filament stiffness $C$ to be particularly important
for controlling not only the structure and properties of self-assembled bundles,
but also the sensitivity of these properties -- size, twist and stability -- to
changes in the availability and affinity of crosslinkers.

The effective torsional elastic energy of bundles derived in Sec. III describes
an intrinsic frustration between shear distortion of crosslinks that bind
together neighboring filament grooves and the torsional elastic response of
filaments themselves.  In the limiting case of perfectly rigid linkers, where
$\lambda \to 0$, our model is not unlike the elastic model of coiled-coils of
Neukirch, Goriely and Hausrath~\cite{neukirch}, in which adhesive interactions
between neighboring helical molecules maintain grooves in perfect registry.
Geometrically, this can be accomplished either by untwisting the rotation of the
grooves or by superhelically twisting the filaments about the bundle axis. The
twist elastic energy of the filaments is minimal for the state of perfect
superhelical twist, $\Omega=\omega_0$. Deviations from this optimal geometry,
while maintaining perfect groove contact, are described by a coarse-grained
twist free-energy cost, $f_{twist} = \rho C(\Omega- \omega_0)^2/2$.

Our theory generalizes the coiled-coil model of perfect groove contact in two
ways.  First, grooves only maintain contact over fraction, $\rho$, of the length
of the filaments due to the finite density of crosslinkers between any filament
pair.  This distinction accounts for the $\rho$ dependence of the effective
torsional modulus, $\rho C$ of the stiff-linker regime, a significant effect in
the context of linker-mediated assembly.  

The second important and novel effect captured by the model is the elastic
compliance of the linker bonds themselves.  Flexibility of bound linkers gives
rise to highly nonlinear torsional properties of the bundles themselves.  
The twist elastic behavior predicted by the case of groove-locked contact, gives
way to a highly-twist compliant state (``groove-slip'') when deviations from the
optimal twist geometry are large, $|\Omega - \omega_0| \gtrsim \delta \Omega_c$.
The crossover twist $\delta\Omega_c\sim \rho/\lambda$, which delineates the
groove-locked from the groove-slip regime, increases both with increased linker density, $\rho$ and increased linker stiffness $\Gamma$.  Hence, untwisting bundles out of the groove-locked into the groove-slip state
indicates a shift in the mechanical load from filament twist to in-plane
crosslink shear distortions. In this limit the elastic cost is insensitive to
bundle twist $\Omega$, and given by $f_{twist} =\rho C\delta\Omega_c^2/2\simeq
\rho^3\Gamma$.

The distinction between a purely harmonic twist energy for stiff linkers and a
non-linear twist energy for flexible linkers has key consequences for the
structural and mechanical properties of crosslinked, helical filament bundles.
This is reflected in the optimal rate of bundle twist, a property that is
determined not only by a balance of the cost of linker shear and filament twist,
but also the mechanical costs associated with filament bend.  The costs of
filament bending are themselves highly sensitive to the lateral radius of the
bundle due to the increased curvature of filaments away from the bundle
center~\cite{grason_pre_09}.  In the case of rigid linkers grooves remain locked
in close contact, so that the optimal twist derives purely from the competition
between the bending cost of rotating filaments superhelically in the bundle, and
the twist elastic cost needed to maintain groove-alignment when filaments are
parallel.  This balance is described by the results of eq. (\ref{eq: Omlocked}),
where the increased cost of bending filaments in large bundles leads to a
continuous unwinding as $\Omega \sim R^{-2/3}$.  

In contrast, bundles bound by relatively flexible linkers are much more
sensitive to bundle size, the sensitivity ultimately becoming singular when $ \delta \Omega_c \leq 0.362 \omega_0$.  These
flexibly crosslinked bundles exhibit a discontinuous drop in $\Omega$ as a
function of $R$, as the bundle rapidly jumps between the groove-locked and
groove-slip behaviors.  Hence, in this flexible linker regime, the unwinding of
bundle twist with increased size is largely determined by a competition between
filament bending and linker shear due to misaligned grooves.  A similar distinction is predicted for the non-linear torsional response of bundles: $\Omega$ depends continuously on externally applied torque for rigidly
crosslinked bundles; while flexibly crosslinked bundles are multi-stable,
exhibiting discontinuous transitions between groove-locked and groove-slip
states driven by external torque.

These non-linear structural and mechanical properties reflect the frustration of
the geometry of crosslinking within helical filament bundles that ultimately
gives rise to intrinsic mechanical torques that lead to {\it self-limiting
  bundle assembly}.  Considering the thermodynamically optimal state of twist
and radius for a system of associating filaments and linkers, we found two basic
scenarios for the dependence of assembly behavior on the number and affinity of
crosslinkers, as parameterized by $\rho$ and $\epsilon$, respectively.  For the
case of {\it rigid linkers} (shown in Fig.~\ref{fig.phasediagram}a), we find
that assembled filaments maintain the high-torque, groove-locked state over the
relevant range of their assembly.  Above a critical cohesive energy per
crosslinker, $\epsilon_c/2$, we predict that filament assemblies form at all
linker densities, with macroscopically large and untwisted assembly behavior at
low $\rho$ crossing over continuously to a regime of twisted and finite sized
bundles at large $\rho$.  For less cohesive crosslinking bonds,
$\epsilon<\epsilon_c/2$, we find that filaments remain largely unbundled below a
linker density $\rho \lesssim \rho_c$, above which they form twisted bundles
whose lateral size grows as $R_* \sim \rho^{1/3}$.  For bundles crosslinked by
{\it flexible linkers} (behavior shown in Fig.~\ref{fig.phasediagram}a) the
transition between highly and weakly twisted states as function of increased
$\rho$ is discontinuous below a critical value of $\epsilon$.  We also find that
finite-sized bundles, untwisted macroscopic assemblies and unbundled, free
filaments coexist at a triple point whose precise location is sensitive to the
flexibility of linkers as parameterized by $\rho_S$.  For values of $\epsilon$
below this triple point, we predict that macroscopic assembly occurs in the
limit of small $\rho$, giving way to a dispersed filament phase at intermediate
$\rho$-values and at the highest range of linker density, a re-entrant phase of
finite-sized linker mediated bundles occurs.

The rich spectrum of possible assembly behavior can be fully classified in terms of three parameters.  The fundamental cohesive
energy scale is determined by $\epsilon_c \propto C \omega_0^2 \sigma^{-1}$,
which corresponds to the energy of untwisting the intrinsic rotation of the helical grooves of the filaments.
The different modes of elastic deformation induced by crosslinkers in helical
filament bundles account for the presence of two fundamental scales of bound
crosslinker fraction, $\rho$.  The first, $\rho_c \propto K (\omega_0 d)^2/C$,
can be understood as the ratio of the mechanical costs of two pair-wise filament
geometries: $K \omega_0^4 d^2$, roughly the bend cost of winding a filament pair
into a groove-locked coiled-coil geometry; and $C \omega_0^2$, the mechanical
cost of untwisting the intrinsic helical grooves into parallel filaments.  We
find that for flexible crosslinks, $\rho_c$ sets a lower limit for the fraction
of bound crosslinkers at which self-limited bundles form, suggesting that
self-assembled bundles are thermodynamically favored when the cost of filament
bend is relatively small compared to the cost of filament twisting.  The final
parameter, $\rho_S \propto \lambda\omega_0\propto (\Gamma / C)^{1/2} \omega_0$,
reflects the relative elastic cost of linker shear to filament twist and
determines the density of crosslinkers below which high-torque, groove-locked
bundle states crossover to low-torque, groove-slip behavior.  Hence, we can
classify the thermodynamic distinctions between linker-mediated assembly by the
ratio $\rho_c/ \rho_S$ which is much greater or less than 1 for the respective
rigid and flexible behaviors depicted in Fig.~\ref{fig.phasediagram}.

The predictions of our coarse-grained model are most directly relevant to the
formation of parallel actin bundles~\cite{revenu}, self-organized cytoskeletal
filament assemblies that form in a variety of cellular specializations under the
influence of compact crosslinking proteins~\cite{pollard, bartles, faix}.  There
is a range of experimental evidence of parallel actin bundles formed {\it in
  vivo}~\cite{tilney_saunders, derosier_censullo, tilney_derosier_mulroy,
  derosier_tilney} and {\it in vitro}~\cite{claessens_pnas_08, purdy, shin}
showing that the presence of certain crosslinking proteins affects the torsional
state of bundled filaments, leading to a modest adjustment of the rotation rate
of primary helix formed by the actin monomers.  Recent small-angle scattering
studies suggest, in fact, that the presence of different crosslinking proteins
modify the twist of bundle actin filaments to a similar degree, but
reconstituted solutions of actin bundles exhibit a remarkably different
sensitivity to the concentration of crosslinking proteins~\cite{shin}.  The
crosslinking protein fascin, a primary component of filopodial
bundles~\cite{faix}, was shown to assemble filaments into bundles with a
continuously variable degree of filament twist, while the crosslinking protein
espin~\cite{bartles}, an abundant crosslinker in microvillar and
mechanosensitive stereosciliar bundles, drove a discrete transition between the
native state of twist and the torsional geometry of the
``fully-bundled'' actin filaments.  Theoretical studies of a model actin
filaments in bulk, parallel arrays attribute this difference in twist
sensitivity  to differences in the {\it stiffness} of the crosslinking bonds
themselves~\cite{shin, shin_10}.

In our model, we show that the presence of crosslinking bonds in parallel arrays
of helical filaments not only modifies the torsional geometry of individual
filaments, but in general determines the amount of global twist of the entire
bundle assembly. In particular, we show that the competition between the inter-filament geometry
preferred by crosslinking and the bend elasticity of filaments mediates an
intrinsic feedback between the lateral assembly size and the amount of
superhelical twist.  An important and general prediction of our model is the appearance of a
thermodynamically stable phase of finite radius bundles at sufficiently high
crosslink fractions that exhibits a linker dependent equilibrium radius, $R_*
\sim \rho^{1/3}$.  Claessens {\it et al} has explored the dependence of actin
bundles formed in reconstituted solutions of filaments and the crosslinker,
fascin~\cite{claessens_pnas_08}.  Above a critical ratio of fascin to actin
monomer, $r$, they found that the mean diameter of bundles formed exhibited a
powerlaw dependence, $R_*\sim r^{0.3}$, for small bundles, consistent with the
predictions our coarse-grained model, which predict that the radius of twisted
bundles grows with bound linker fractions as $R_*\sim \rho^{1/3}$.  Hence, our
model establishes a thermodynamic link between simultaneous influence of
crosslinker fraction on bundle size and degree of filament twist in these
actin/fascin experiments.  Additionally, our theory establishes a range of
further predictions on both the affinity and flexibility of crosslinking
proteins, suggesting the need for further experiments on the influence of size
and structure of parallel actin bundles on properties of crosslinkers.  Notably,
the implied differences in the compliance between fascin and espin crosslinks
should be correlated with measurably differences in the assembly thermodynamics
of reconstituted bundle forming solutions.

In summary, we have found that although crosslinking bonds between helical
filaments mediate net cohesive interactions between filaments, the complex
interplay between the mechanics of linker and filament distortions ultimately
gives rise to a nontrivial dependence of the structure of self-assembled bundles
on the number of bound linkers.  We quantitatively model the thermodynamic
influence of a number of microscopic parameters, from filament stiffness and
intrinsic twist to the affinity of crosslink bonds.  Among these, we show that
the stiffness of the linkers themselves account for remarkable differences in
the sensitivity of bundle structure to the availability of crosslinks.  It is
natural to expect that living systems may exploit the intrinsic frustration of
crosslinking in helical filament bundles as robust means of regulating the size
and structure of self-assembled bundles, not by modifying properties of the
filaments, but instead by carefully regulating the number and type of
crosslinking proteins alone.

\begin{acknowledgments}
  The authors would like to acknowledge H. Shin for useful comments, and the
  hospitality of the Aspen Center for Physics, where this study originated.  CH was
  supported by a Feodor Lynen fellowship of the German Humboldt Foundation.
 GG was supported by the NSF Career program under DMR Grant 09-55760. \end{acknowledgments}

\begin{appendix}
\section{Out-of-plane shear}

In this appendix we demonstrate how out-of-plane shear deformations of
crosslinkers may partially relax through a coupling to in-plane shear modes.
Generalizing eq.(\ref{eq: elink}) we consider an energy of crosslinking,
\begin{equation}
  \label{eq: elink.appendix}
  E_{link} = - \epsilon + \frac{k_\perp}{2} \Delta_\perp^2 +\frac{k_\parallel}{2} \Delta_\parallel^2\,,
\end{equation}
that contains, next to the in-plane shear deformations, $\Delta_\perp$, the
out-of plane shear, $\Delta_\parallel$, which can be associated with filaments
tilting into the plane of lattice order.
 In a twisted bundle, filaments are increasingly tilted along the azimuthal
direction as the radial distance from the bundle center increases,
\begin{equation}
\tv(\rv) \simeq \hat{z} + \Omega \hat{z} \times \rv,
\end{equation}
where $\rv$ is the radial vector from the center of the bundle.  The in-plane
distance between crosslinking points is fixed to $a'=d-2a$, and we can relate
the out-of-plane shear to the average tilt angle, $\theta_{ij}$, of two
neighboring filaments 
\begin{equation}
\label{eq: dpar}
\Delta_\parallel = 2 a' \sin \theta_{ij} = 2 a' \tv \cdot \hat{\rv}_{ij} = 2 a' \Omega ( \hat{z} \times \rv) \cdot \hat{\rv}_{ij}
\end{equation}
where $\rv$ in this equation refers to the mean in-plane position of filaments
$i$ and $j$ and $\hat{\rv}_{ij}$ is a unit vector that points from one filament
to the other. Hence, in the twisted geometry, out-of-plane shear depends on
$|\hat{\rv}_{ij}\times \rv|$, and is different for different filament pairs in
the bundle.  As we are interested in $\Delta_\parallel^2$, we can replace
$|\hat{\rv}_{ij}\times \rv|^2$ with its average over the 6-fold nearest neighbor
directions on the hexagonal lattice, $ \langle |\hat{\rv}_{ij}\times \rv|^2
\rangle = r^2 /2$ and the neighbor-averaged shear becomes,
\begin{equation}
\langle \Delta_\parallel^2 \rangle = 2 a'^2 (\Omega r)^2 .
\end{equation}
Notice that this shear is insensitive to $\phi(z)$, unlike $\Delta_\perp$.

As described in eq. (\ref{eq: dpar}), out-of-plane shearing occurs when
neighboring filaments, say $+$ and $-$, sharing crosslinks, tilt along their
nearest neighbor separation vector, $\hat{\rv}_{+ -}$, making an angle
$\theta_0$, relative to the straight configuration as shown in Fig.~\ref{fig:
  outofplane}.

This assumes the crosslinks to bind in the ``horizontal'' $x-y$ plane, which is
perpendicular to the bundle axis. However, if we allow the crosslinks to relax
their binding geometry at fixed $\rho$ after shear takes place, then the actual
shear angle of the crosslinks $\theta$, may be reduced from the shear angle of
the crosslinks, $\theta_0$. This proceeds by shifting the linker ends bound to the $+$ ($-$) filament
``upstream'' (``downstream''), by an amount $\delta z_+ = \delta/2$ ($\delta z_- = \delta/2$) and reduces the out of
plane shear to $\Delta_\parallel = a ' \theta_0 - \delta$.

\begin{figure}
  \center \includegraphics[width = 2.5in]{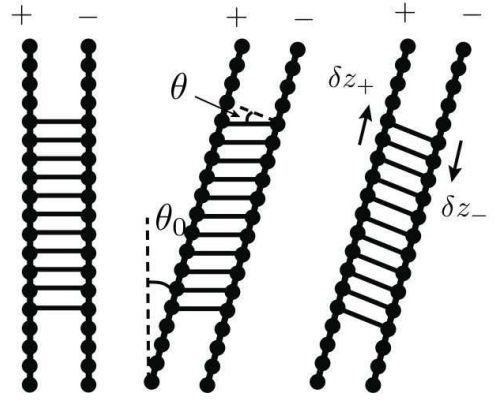}\caption{A schematic of
    linker binding reorganization in response to an out-of-plane shear cost for
    2 filaments grooves labeled ``+" and ``-". Linkers are sheared with
    respect to the filament backbones by angle, $\theta$, when a filament pair
    tilts along the direction of nearest neighbor separation, as shown in the
    middle.  The shifting the bound ends of linkers up- and down-stream
    respectively from the + and - filaments, the links may relax the shear angle
    made with respect to the filament from the tilt angle, $\theta_{+-}$, as
    shown on the right.}
\label{fig: outofplane}
\end{figure}

The amount of linker shift, $\delta$, is determined from the competing effect of
in-plane shear. Due to the rotation of the filament groove, as described by
$\psi'$, the shift of crosslink ends {\it also} alters the in-plane shear angle
to $\psi+ \psi' \delta$, where we assume that the rotation profile for both
filaments is identical.  Integrating the in-plane shear along one binding zone
from $z = - \rho \ell$ to $z = +\rho \ell$ and noting from the analysis of
Sec.\ref{sec:elastic-free-energy} that $\psi(z)=-\psi(z)$ while $\psi'(z) = \psi'(-z)$, the
elastic contribution per unit length from out of plane shear from a
single-binding zone can be written as,
\begin{equation}
  E_\parallel (\delta)/L= \frac{\Gamma \rho}{8} \delta^2 \langle \psi
  '^2\rangle_\rho + \frac{\rho \Gamma'}{2} \Big( \theta_0 - \frac{\delta}{a'}
  \Big)^2, 
\end{equation}
where
\begin{equation}
  \langle \psi '^2\rangle_\rho \equiv \frac{\int_{-\rho \ell}^{\rho \ell} dz ~ \psi'(z) }{\rho \ell} ,
\end{equation}
and $\Gamma' = k_\parallel a'^2 \sigma$.  Minimizing over the shift of bound
linker ends we find,
\begin{equation}
  \delta_* / a' = \theta_0 \frac{ \Gamma'}{ \Gamma' + \Gamma a'^2/ 4} ,
\end{equation}
and an effective free energy contribution per unit length of the interfilament
groove,
\begin{equation}
  \label{eq: epar}
  E_\parallel /L = \frac{\rho}{2} \theta_0^2 \langle \psi '^2\rangle_\rho \frac{
    \Gamma \Gamma' a'^2 }{4 \Gamma' + \Gamma  \langle \psi '^2\rangle_\rho  a'^2
  } . 
\end{equation}
Thus, we find that the coupling between out-of-plane and in-plane shear modes
leads to a tilt-dependence renormalization of the torsional energy of the
filaments ($\sim C\langle \psi '^2\rangle_\rho$).

In the limit that linkers are softer to out-of-plane shear than to in-plane
shear, or $\Gamma' \ll \Gamma$, the contribution to the bundle energy from the
out-of-plane shear is clear negligible compared to the in-plane cost. From eq.
(\ref{eq: epar}) we find that also in the limit that $\Gamma' \gg \Gamma$, the
out-of-plane shear leads to a renormalization of the effective twist modulus of
filaments, $C_{eff} = C + \rho (\Omega r)^2 \Gamma a'^2 /16$, using $\langle
\theta_0^2\rangle = (\Omega r)^2/2$.  Relative to the effective energy analyzed
in Sec.\ref{sec:prop-bundl-with} and \ref{sec:link-medi-filam}, the net correction to the
linker mediated-twist energy of the bundle from out-of-plane shear mode
ultimately represents a higher-order term in $\Omega$, proportional to
$\Omega^2(\Omega - \omega_0)^2$, and therefore, will not strongly alter the
quantitative analysis of bundle thermodynamics from the $\Gamma' = 0$ case.

\end{appendix}


\begin{thebibliography}{10}

\bibitem{fratzl} P. Fratzl, Cur. Opin. Coll. Int. Sci. {\bf 8}, 32 (2003).

\bibitem{theriot} S. M. Rafelski and J. A. Theriot, Ann. Rev. Biochem. {\bf 73}, 209 (2004).

\bibitem{revenu} C. Revenu, R. Athman, S. Robine, and D. Louvard,  Nat. Rev. Mole. Cell Biol. {\bf 5}, 635 (2004).

\bibitem{pollard} T. D. Pollard and J. A. Cooper,  Ann. Rev. Biochem. {\bf 55}, 987 (1986).

\bibitem{bartles} J. R. Bartles,  Curr. Opin. Cell. Biol. {\bf 2}, 72 (2000).


\bibitem{faix} J. Faix and K. Rottner,  Curr. Opin. Cell Biol. {\bf 18}, 18 (2006).

\bibitem{aratyn} Y. S. Aratyn, T. E. Schaus, E. W. Taylor, and G. G. Borisy, Mol. Biol. Cell. {\bf 18}, 3928 (2007).



\bibitem{shin_mahadevan} J. H. Shin, L. Mahadevan, P. T. So and P. Matsudaira, J. Mol. Biol. {\bf 337} 255  (2004).


\bibitem{shin_mahadevan_pnas} J. H. Shin, M. Gardel, L. Mahadevan, P. Matsudaira and D. A. Weitz, Proc. Nat. Acad. Sci. USA {\bf 101} 9636  (2004).



\bibitem{claessens_nat_06} M. M. A. E. Claessens, M. Bathe, E. Frey, and A. R. Bausch, Nat. Materials {\bf 5}, 748  (2008).

\bibitem{claessens_pnas_08} M. M. A. E. Claessens, C. Semmrich, L. Ramos, and A. R. Bausch, Proc. Natl. Acad. Sci. USA {\bf 105}, 8819 (2008).


\bibitem{purdy} K. R. Purdy, J. R. Bartles, and G. C. L. Wong. 2007,  Phys. Rev. Lett. {\bf 98} 058105 (2007).

\bibitem{shin}  H. Shin, K. R. Purdy Drew, J. R. Bartles, G. C. L. Wong, and G. M. Grason, Phys. Rev. Lett. {\bf 103} 238102 (2009).

\bibitem{squire}
J. M. Squire and P. J. Vibert, eds. {\it Fibrous Protein Structure} (Academic Press, London, 1987).

\bibitem{derosier_censullo} D. J. DeRosier and R. Censullo, J. Mol. Biol. {\bf 146}, 77 (1981).


\bibitem{tilney_derosier_mulroy} L. G. Tilney, D. J. DeRosier and M. J. Mulroy, J. Cell Biol.  {\bf 86}, 244 (1980).  

\bibitem{derosier_tilney} D. J. DeRosier and L. G. Tilney, Cold Spring Harb. Symp. Quant. Biol. {\bf 46}, 525 (1982).

\bibitem{tilney_saunders} L. G. Tilney, E. H. Egelman, D. J. DeRosier and J. C. Saunders, J. Cell. Biol.  {\bf 96}, 822 (1983).


\bibitem{weisel}
J. W. Weisel, C. Nagaswami and L. Makowski, Proc. Nat. Acad. Sci. USA {\bf 84}, 8991 (1987).

\bibitem{ottani}
V. Ottani, D. Martinin, M. Franchi, A. Ruggeri and M. Raspanti, Micron {\bf 33}, 587 (2002).


\bibitem{turner} M. S. Turner, R. W. Briehl, F. A. Ferrone and R. Josephs, Phys. Rev. Lett. {\bf 90}, 128103 (2003).

\bibitem{grason_prl_07} G. M. Grason and R. F. Bruinsma, Phys. Rev. Lett. {\bf 99}, 098101 (2007).

\bibitem{grason_pre_09} G. M. Grason,  Phys. Rev. E  {\bf 79}, 041919 (2009).

\bibitem{hagan} Y. Yang, R. B. Meyer and M. F. Hagan, Phys. Rev. Lett.  {\bf 104}, 258102 (2010).

\bibitem{bathe} M. Bathe, C. Heussinger, M.M.A.E Claessens, A. Bausch and
  E.  Frey, Biophys. J. {\bf 94}, 2955 (2008).

\bibitem{heussinger} C. Heussinger, M. Bathe, E. Frey, Phys. Rev. Lett. {\bf 99}, 048101 (2007).

\bibitem{heussinger_pre}
C. Heussinger, F. Sch\"uller and E. Frey, Phys. Rev. E {\bf 81}, 021904 (2010).

\bibitem{neukirch} S. Neukirch, A. Goriely and A. C. Hausrath, Phys. Rev. Lett. {\bf 100}, 038105 (2008).


\bibitem{holmes} K. Holmes, D. Popp, W. Gebhard and W. Kabsch, Nature {\bf 347}, 44(1990).  

\bibitem{shin_10} H. Shin and G. M. Grason, Phys. Rev. E {\bf 82}, 051919 (2010).

\bibitem{wolgemuth_sun} O. N. Yogurtcu, C. W. Wolgemuth and S. X. Sun, Biophys.
  J. {\bf 99}, 3892 (2010).

\bibitem{nelson} P. Nelson, Biophys. J. {\bf 74}, 2501 (1998).

\bibitem{strehle} D. Strehle {\it et al.}, Eur. Biophys. J. {\bf 40}, 93 (2011).










\end{thebibliography}
\end{document}